\documentstyle[12pt,epsf]{article}
\newcommand{\nc}{\newcommand}
\nc{\renc}{\renewcommand}

\renc{\baselinestretch}{1.2}
\nc{\com}[1]{\\{\bf \# {#1}}\\}
\newlength{\overeqskip}
\newlength{\undereqskip}
\nc{\be}[1]{\begin{equation} \mbox{$\label{#1}$}}
\nc{\bea}[1]{\begin{eqnarray} \mbox{$\label{#1}$}}
\nc{\Section}[2]{\section{\sc #2}\label{#1}}
\nc{\Bibitem}[1]{\bibitem{#1}}
\nc{\Label}[1]{\label{#1}}

\nc{\eea}{\vspace{\undereqskip}\end{eqnarray}}
\nc{\ee}{\vspace{\undereqskip}\end{equation}}
\nc{\bdm}{\begin{displaymath}}
\nc{\edm}{\end{displaymath}}
\nc{\dpsty}{\displaystyle}
\nc{\bc}{\begin{center}}
\nc{\ec}{\end{center}}
\nc{\ba}{\begin{array}}
\nc{\ea}{\end{array}}
\nc{\bab}{\begin{abstract}}
\nc{\eab}{\end{abstract}}
\nc{\btab}{\begin{tabular}}
\nc{\etab}{\end{tabular}}
\nc{\bit}{\begin{itemize}}
\nc{\eit}{\end{itemize}}
\nc{\ben}{\begin{enumerate}}
\nc{\een}{\end{enumerate}}
\nc{\bfig}{\begin{figure}}
\nc{\efig}{\end{figure}}
\nc{\eqs}[2]{\mbox{Eqs.~(\ref{#1},\,\ref{#2})}}
\nc{\eq}[1]{\mbox{Eq.~(\ref{#1})}}

\nc{\figs}[2]{\mbox{Figs.(\ref{#1},\,\ref{#2})}}
\nc{\fig}[1]{\mbox{Fig.(\ref{#1})}}

\nc{\figcap}[1]{\refstepcounter{figure}
	{\bf Figure \thefigure}: {\small\sl #1}}
\nc{\tabcap}[1]{\refstepcounter{table}
	{\bf Table \thetable}: {\small\sl #1}}

\nc{\tag}[1]{\label{#1} \marginpar{{\footnotesize #1}}}
\nc{\mtag}[1]{\label{#1} \mbox{\marginpar{{\footnotesize #1}}}}

\nc{\etal}{\mbox{\it et al. }}
\nc{\ie}{{\it i.e.}}
\nc{\eg}{{\it e.g.}}

\nc{\arreq}{&\!\!=\!\!&}
\nc{\arrmi}{&\!\!-\!\!&}
\nc{\arrpl}{&\!\!+\!\!&}
\nc{\arrap}{&\!\!\!\approx\!\!\!&}
\nc{\nn}{\nonumber\\}
\nc{\align}{\!\!\!\!\!\!\!\!&&}

\def\lsim{\; \raise0.3ex\hbox{$<$\kern-0.75em
      \raise-1.1ex\hbox{$\sim$}}\; }
\def\gsim{\; \raise0.3ex\hbox{$>$\kern-0.75em
      \raise-1.1ex\hbox{$\sim$}}\; }
\nc{\DOT}{\hspace{-0.08in}{\bf .}\hspace{0.1in}}
\nc{\Laada}{\hbox {$\sqcap$ \kern -1em $\sqcup$}}
\nc\loota{{\scriptstyle\sqcap\kern-0.55em\hbox{$\scriptstyle\sqcup$}}}
\nc\Loota{{\sqcap\kern-0.65em\hbox{$\sqcup$}}}
\nc\laada{\Loota}
\nc{\qed}{\hskip 3em \hbox{\BOX} \vskip 2ex}
\def\Re{{\rm Re}\hskip2pt}

\nc{\real}{{\rm I \! R}}
\nc{\Z}{{\sf Z \!\!\! Z}}
\nc{\complex}{{\rm C\!\!\! {\sf I}\,\,}}
\def\bigid{\leavevmode\hbox{\small1\kern-3.8pt\normalsize1}}
\def\id{\leavevmode\hbox{\small1\kern-3.3pt\normalsize1}}
\nc{\slask}{\!\!\!/}
\nc{\bis}{{\prime\prime}}
\nc{\pa}{\partial}
\nc{\na}{\nabla}
\nc{\ra}{\rangle}
\nc{\la}{\langle}
\nc{\goto}{\rightarrow}
\nc{\swap}{\leftrightarrow}

\nc{\EE}[1]{ \mbox{$\cdot10^{#1}$} }
\nc{\abs}[1]{\left|#1\right|}
\nc{\at}[2]{\left.#1\right|_{#2}}
\nc{\norm}[1]{\|#1\|}
\nc{\abscut}[2]{\Abs{#1}_{\scriptscriptstyle#2}}
\nc{\vek}[1]{{\rm\bf #1}}
\nc{\integral}[2]{\int\limits_{#1}^{#2}}
\nc{\inv}[1]{\frac{1}{#1}}
\nc{\dd}[2]{{{\partial #1}\over{\partial #2}}}
\nc{\ddd}[2]{{{{\partial}^2 #1}\over{\partial {#2}^2}}}
\nc{\dddd}[3]{{{{\partial}^2 #1}\over
	{\partial #2 \partial #3}}}
\nc{\dder}[2]{{{d #1}\over{d #2}}}
\nc{\ddder}[2]{{{d^2 #1}\over{d {#2}^2}}}
\nc{\dddder}[3]{{d^2 #1}\over
	{d #2 d #3}}
\nc{\dx}[1]{d\,^{#1}x}
\nc{\dy}[1]{d\,^{#1}y}
\nc{\dz}[1]{d\,^{#1}z}
\nc{\dl}[1]{\frac{d\,^{#1}l}{(2\pi)^{#1}}}
\nc{\dk}[1]{\frac{d\,^{#1}k}{(2\pi)^{#1}}}
\nc{\dq}[1]{\frac{d\,^{#1}q}{(2\pi)^{#1}}}

\nc{\cc}{\mbox{$c.c.$ }}
\nc{\hc}{\mbox{$h.c.$ }}
\nc{\cf}{cf.\ }
\nc{\erfc}{{\rm erfc}}
\nc{\Tr}{{\rm Tr\,}}
\nc{\tr}{{\rm tr\,}}
\nc{\pol}{{\rm pol}}
\nc{\sign}{{\rm sign}}
\nc{\bfT}{{\bf T }}

\nc{\cA}{{\cal A}}
\nc{\cB}{{\cal B}}
\nc{\cD}{{\cal D}}
\nc{\cE}{{\cal E}}
\nc{\cF}{{\cal F}}
\nc{\cG}{{\cal G}}
\nc{\cH}{{\cal H}}
\nc{\cL}{{\cal L}}
\nc{\cO}{{\cal O}}
\nc{\cT}{{\cal T}}
\nc{\rvac}[1]{|{\cal O}#1\rangle}
\nc{\lvac}[1]{\langle{\cal O}#1|}
\nc{\rvacb}[1]{|{\cal O}_\beta #1\rangle}
\nc{\lvacb}[1]{\langle{\cal O}_\beta #1 |}
\nc{\bb}{\bar{\beta}}
\nc{\ctH}{\tilde{\cal H}}
\nc{\chH}{\hat{\cal H}}
\nc{\1}{\aa}
\nc{\2}{\"{a}}
\nc{\3}{\"{o}}
\nc{\4}{\AA}
\nc{\5}{\"{A}}
\nc{\6}{\"{O}}
\nc{\al}{\alpha}
\nc{\Del}{\Delta}
\nc{\e}{\epsilon}
\nc{\eps}{\epsilon}
\nc{\lam}{\lambda}
\nc{\om}{\omega}
\nc{\Om}{\Omega}
\nc{\ve}{\varepsilon}
\nc{\mn}{{\mu\nu}}
\nc{\k}{\kappa}
\nc{\vp}{\varphi}

\nc{\advp}[3]{{\it  Adv.\ in\ Phys.\ }{{\bf #1} {(#2)} {#3}}}
\nc{\annp}[3]{{\it  Ann.\ Phys.\ (N.Y.)\ }{{\bf #1} {(#2)} {#3}}}
\nc{\apl}[3]{{\it  Appl. Phys. Lett. }{{\bf #1} {(#2)} {#3}}}
\nc{\apj}[3]{{\it  Ap.\ J.\ }{{\bf #1} {(#2)} {#3}}}
\nc{\apjl}[3]{{\it  Ap.\ J.\ Lett.\ }{{\bf #1} {(#2)} {#3}}}
\nc{\app}[3]{{\it Astropart.\ Phys.\ }{{\bf #1} {(#2)} {#3}}}
\nc{\cmp}[3]{{\it  Comm.\ Math.\ Phys.\ }{{ \bf #1} {(#2)} {#3}}}
\nc{\cqg}[3]{{\it  Class.\ Quant.\ Grav.\ }{{\bf #1} {(#2)} {#3}}}
\nc{\epl}[3]{{\it  Europhys.\ Lett.\ }{{\bf #1} {(#2)} {#3}}}
\nc{\ijmp}[3]{{\it Int.\ J.\ Mod.\ Phys.\ }{{\bf #1} {(#2)} {#3}}}
\nc{\ijtp}[3]{{\it Int.\ J.\ Theor.\ Phys.\ }{{\bf #1} {(#2)} {#3}}}
\nc{\jmp}[3]{{\it  J.\ Math.\ Phys.\ }{{ \bf #1} {(#2)} {#3}}}
\nc{\jpa}[3]{{\it  J.\ Phys.\ A\ }{{\bf #1} {(#2)} {#3}}}
\nc{\jpc}[3]{{\it  J.\ Phys.\ C\ }{{\bf #1} {(#2)} {#3}}}
\nc{\jap}[3]{{\it J.\ Appl.\ Phys.\ }{{\bf #1} {(#2)} {#3}}}
\nc{\jpsj}[3]{{\it J.\ Phys.\ Soc.\ Japan\ }{{\bf #1} {(#2)} {#3}}}
\nc{\lmp}[3]{{\it Lett.\ Math.\ Phys.\ }{{\bf #1} {(#2)} {#3}}}
\nc{\mpl}[3]{{\it  Mod.\ Phys.\ Lett.\ }{{\bf #1} {(#2)} {#3}}}
\nc{\ncim}[3]{{\it  Nuov.\ Cim.\ }{{\bf #1} {(#2)} {#3}}}
\nc{\np}[3]{{\it  Nucl.\ Phys.\ }{{\bf #1} {(#2)} {#3}}}
\nc{\pr}[3]{{\it Phys.\ Rev.\ }{{\bf #1} {(#2)} {#3}}}
\nc{\pra}[3]{{\it  Phys.\ Rev.\ A\ }{{\bf #1} {(#2)} {#3}}}
\nc{\prb}[3]{{\it  Phys.\ Rev.\ B\ }{{{\bf #1} {(#2)} {#3}}}}
\nc{\prc}[3]{{\it  Phys.\ Rev.\ C\ }{{\bf #1} {(#2)} {#3}}}
\nc{\prd}[3]{{\it  Phys.\ Rev.\ D\ }{{\bf #1} {(#2)} {#3}}}
\nc{\prl}[3]{{\it Phys.\ Rev.\ Lett.\ }{{\bf #1} {(#2)} {#3}}}
\nc{\pl}[3]{{\it  Phys.\ Lett.\ }{{\bf #1} {(#2)} {#3}}}
\nc{\prep}[3]{{\it Phys\. Rep.\ }{{\bf #1} {(#2)} {#3}}}
\nc{\prsl}[3]{{\it Proc.\ R.\ Soc.\ London\ }{{\bf #1} {(#2)} {#3}}}
\nc{\ptp}[3]{{\it  Prog.\ Theor.\ Phys.\ }{{\bf #1} {(#2)} {#3}}}
\nc{\ptps}[3]{{\it  Prog\ Theor.\ Phys.\ suppl.\ }{{\bf #1} {(#2)} {#3}}}
\nc{\physa}[3]{{\it  Physica\ A\ }{{\bf #1} {(#2)} {#3}}}
\nc{\physb}[3]{{\it  Physica\ B\ }{{\bf #1} {(#2)} {#3}}}
\nc{\phys}[3]{{\it Physica\ }{{\bf #1} {(#2)} {#3}}}
\nc{\rmp}[3]{{\it  Rev.\ Mod.\ Phys.\ }{{\bf #1} {(#2)} {#3}}}
\nc{\rpp}[3]{{\it Rep.\ Prog.\ Phys.\ }{{\bf #1} {(#2)} {#3}}}
\nc{\sjnp}[3]{{\it Sov.\ J.\ Nucl.\ Phys.\ }{{\bf #1} {(#2)} {#3}}}
\nc{\spjetp}[3]{{\it Sov.\ Phys.\ JETP\ }{{\bf #1} {(#2)} {#3}}}
\nc{\yf}[3]{{\it Yad.\ Fiz.\ }{{\bf #1} {(#2)} {#3}}}
\nc{\zetp}[3]{{\it Zh.\ Eksp.\ Teor.\ Fiz.\ }{{\bf #1} {(#2)} {#3}}}
\nc{\zp}[3]{{\it Z.\ Phys.\ }{{\bf #1} {(#2)} {#3}}}
\nc{\ibid}[3]{{\sl ibid.\ }{{\bf #1} {#2} {#3}}}
\nc{\rf}[1]{(\ref{#1})}
\nc{\Lett}{\cL^{(1)}(x)}
\nc{\hx}{\hat{x}}
\nc{\hp}{\hat{p}}
\nc{\sr}{s\sqrt{2\omega}}
\nc{\bk}{\beta_\kappa}
\nc{\g}{g(T)}
\nc{\as}{\alpha_s}
\begin{document}
%
\large
\thispagestyle{empty}
\begin{flushright}
	CERN-TH/95-66\\
        hep-ph/9503171
\end{flushright}
\bc
{\Huge\bf Derivatives as an IR Regulator\\[4mm]
 for Massless Fields}
\\[1cm]
\ec
\vspace*{1cm}
\bc
{\large{\bf Per Elmfors\footnote{
E-mail: per.elmfors@cern.ch}}\\[5mm]
{\normalsize CERN,
TH Division,
CH-1211 Gen\`{e}ve 23,
Switzerland\\}}
\ec
\vspace*{2cm}
\bc
{\bf Abstract} \\
\ec
{\normalsize
\begin{quotation}
\noindent
The free propagator for the scalar
$\lambda \phi^4$--theory is calculated
exactly up to the second
derivative of a background field.
Using this propagator I compute the one--loop
effective action,
which then contains all powers of the field but with at most
two derivatives acting on each field. The standard derivative
expansion, which only has a finite number of derivatives
in each term, breaks
down for small fields when the mass is zero, while the expression
obtained here has a well--defined expansion in $\phi$. In this way
the resummation of derivatives cures the naive IR divergence.
The extension to finite temperature is also discussed.
\end{quotation}}
\vfill
\newpage
%
\normalsize
\setcounter{page}{1}
\Section{intro}{Introduction}
There are a number of methods for computing
approximations to the effective action depending
on which parameters can be considered as
small. If the field amplitude is small it may
be enough to compute a small
number of $n$-point functions, with the advantage
that they can be computed to all orders in the
derivatives. For instance, the two--point function
can be calculated to all orders in momentum,
yielding corrections to the dispersion relation.
For configurations with large fields all $n$--point
functions become important and have to be resummed.
This we can do for a constant field, thus assuming
all derivatives to be zero, and we get the
effective potential. Even though the effective potential
is calculated for a strictly constant field,
the effective action, expanded to the zeroth order in derivatives,
is obtained from it by simply replacing
the constant by a slowly varying
$x$--dependent field.
By doing so it is assumed that the field varies so slowly that
it can locally be approximated by a constant. Effects that
have to do with global properties,
such as phase separation in spontaneously broken theories
or instanton contributions,
are completely neglected in the quantum corrections,
although there is no direct assumption
that the amplitude of the field variation must be small.
In a derivative expansion of the effective action the next
step is to calculate the wave function renormalization to
all orders in the field without derivatives.
The reason why the effective potential and the wave function
renormalization can be calculated
to all orders in the field from a one--loop calculation is
that the propagator can be constructed exactly in the presence
of a constant background field.  It is, therefore, interesting to see
if one can go further in the local approximation while still
being able to find the exact propagator.
In this paper I use Schwinger's
proper--time method \cite{Schwinger51}
to compute the propagator exactly up
to the second derivative of the field and use this in turn to
compute the effective action and the effective equations of
motion.

I restrict  the calculations here to a toy model, the
$\lambda \phi^4$--model, to study general properties of the method.
It would be interesting to extend the method
to more realistic
theories, in particular gauge theories.
The effective potential has been used
to a large extent in the electroweak model to calculate
bubble properties at the phase transition. The wave
function renormalization constant $Z(\phi)$
has also been computed \cite{JungnickelW94}. There are
two main problems with that procedure.
First, it is very sensitive to the
choice of gauge fixing parameter, because
the usual effective action suffers from
gauge dependence away from its stationary points.
(The Vilkovisky--DeWitt effective action might be the
correct way to get around this problem.)
Secondly, $Z(\phi)$ diverges
when the field goes to zero because the transverse gauge field is
unscreened at high temperature.
In this paper I discuss a way to cure the IR sensitivity
by including higher derivative terms. The idea is that if the
field is much smaller than the scale set by the derivative,
then the derivative expansion does no longer hold and higher
derivatives have to be included exactly. On the other hand,
when the derivatives
are non--negligible, the field cannot be small in any sizeable
volume and in this way the derivatives regulate the IR
sensitivity.
The approximation is still local and there is no hope
that it would cure
the imaginary parts that occur in a spontaneously
broken theory when the mass squared is negative.
It may, therefore, be difficult to apply the method
directly to bubble wall calculations in the electroweak
theory.
%
\Section{Locexp}{Local expansion of the effective action}
The basic equation to use when calculating the effective action
is the generalized tadpole equation \cite{ElmforsEV93}
\be{tadpole}
	\frac{\delta\Gamma[\phi(y)]}
	{\delta\phi(x)}=\Gamma^{(1)}[\phi(y);x]\ ,
\ee
which gives the relation between the effective action $\Gamma[\phi]$
and the one--point function $\Gamma^{(1)}[\phi;x]$
computed in the background of $\phi(y)$.
To find $\Gamma[\phi]$ one has to integrate $\Gamma^{(1)}[\phi;x]$
with respect to $\phi(x)$, but the equation of motion is given
directly by $\Gamma^{(1)}[\phi;x]=0$.
The Feynman rules needed to compute the one--point function
are derived from the Lagrangian in the non--trivial
background
\bea{Lag}
	\cL(\phi)&=&\inv{2}\left((\pa\phi)^2-m^2\phi^2\right)-
	\frac{\lambda}{4!}\phi^4\ , \\
	\cL(\phi+\eta)&=&\cL(\phi)+
	\frac{\delta\Gamma_{cl}}{\delta\phi} \nn
	&&+\inv{2}\left((\pa\eta)^2-(m^2+\frac{\lambda\phi^2}{2})
	\eta^2\right)-\frac{\lambda}{3!}\eta^3-\frac{\lambda}{4!}
	\eta^4\ .
\eea
The tadpole equation to one loop is  then written as
\be{1ltp}
	i\Gamma^{(1)}[\phi;x]=
	i\frac{\delta\Gamma_{cl}}{\delta\phi}+
	\inv{2}(-i\lambda\phi(x))\la x|
	\frac{i}{\hp^2-M^2(\phi)+i\epsilon}|x\ra\ ,
\ee
with $M^2(\phi)=m^2+\lambda\phi^2(\hx)/2$. This equation can be
integrated to give
\be{EA}
	\Gamma[\phi]=\Gamma_{cl}+\frac{i}{2}
	\Tr_x\ln(\hp^2-M^2+i\epsilon)\ .
\ee
Because of the trace over one--particle
states labelled by $x$, there is no ordering
problem when doing the integration, even though $\hp$ and $\phi(\hx)$
do not commute. From \eq{EA} we identify the one--loop Lagrangian
and use the proper--time representation to write it as
\be{1lLag}
	\Lett=-\frac{i}{2}\int_0^\infty\frac{ds}{s}
	\la x|e^{is(\hp^2-M^2+i\epsilon)}|x\ra
	\equiv-\frac{i}{2}\int_0^\infty\frac{ds}{s}
	\la x(s)|x(0)\ra\ .
\ee
This expression has, of course, the usual UV infinities which we
have to renormalize at the end.

In a general background it is not possible to
calculate $\la x|\exp[is(\hp^2-M^2+i\epsilon)]|x\ra$
exactly since $[\hp^2,M^2(\phi(\hx))]\neq 0$. If we think of \eq{1lLag}
as a purely quantum mechanical problem, the non--trivial part of
the amplitude from $|x\ra$ to $\la x|$ comes from  quantum
fluctuations. That is, the virtual particle that propagates
from $x$ back to $x$ probes a certain neighbourghood of $x$
related to the Compton wavelength.
Expanding the background field, i.e. $M^2(\phi)$,
in derivatives should lead to a good approximation if these
fluctuations are not too large. Notice that this is different
from a derivative expansion of
the amplitude itself, which is a complicated function
of $M^2$.
To second order in derivatives we define
\bea{Fom}
	F_\mu&=&\pa_\mu M^2=\lambda\phi\pa_\mu\phi\ , \nn
	\omega_{\mu\nu}&=&\pa_\mu\pa_\nu M^2=\lambda\pa_\mu\phi
	\pa_\nu\phi+\lambda\phi\pa_\mu\pa_\nu\phi\ .
\eea
When computing the Lagrangian at a point $x$ we expand the
position operator
in $M^2(\phi)$ around that point
($x\mapsto x+\hx,\ \hx|x\ra=0$), and define the corresponding
bilinear Hamiltonian
(suppressing the $+i\epsilon$ term in the sequel)
\bea{ham}
	M^2(\phi(x+\hx))&\simeq&M^2+\hx^\mu F_\mu
	+\inv{2}\hx^\mu\omega_{\mu\nu}\hx^\nu\ , \nn
	H&=&\hp^2-M^2-\hx\, F-\inv{2} \hx\, \omega\, \hx\ ,
\eea
using a matrix notation in the second line of \eq{ham}. Following
Schwinger \cite{Schwinger51} the problem has now been reduced
to solving the  differential equation
\bea{de}
	\frac{d}{ds}\la x''(s)|x'(0)\ra &=&
	i\la x''(s)|H|x'(0)\ra\ , \nn
	\la x''(0)|x'(0)\ra &=& \delta(x''-x')\ .
\eea
In order to write \eq{de} as an explicit differential equation,
$H(\hp(0),\hx(0))$ has to be rewritten  as a function of
$\hx(s)$ and $\hx(0)$ using the equations of motion for
$\hp$ and $\hx$
\be{xpeqom}
	\frac{d}{ds}\left( \ba{c} \hx(s) \\ \hp(s) \ea \right)
	=-i\biggl[ H(\hp(s),\hx(s)),
	\left( \ba{c} \hx(s) \\ \hp(s) \ea \right) \bigg]
	=\left(\ba{cc} 0 & 2 \\ \omega & 0 \ea \right)
	\left( \ba{c} \hx(s) \\ \hp(s) \ea \right)
	+\left( \ba{c} 0 \\ F \ea \right)\ .
\ee
The solution is
\be{sol}
	\left(\ba{c} \hx(s)+\omega^{-1}F\\
	\sqrt{\frac{2}{\omega}} \hp(s) \ea\right)
	= \left(\ba{cc} \cosh(s\sqrt{2\omega}) & \sinh(s\sqrt{2\omega})\\
	\sinh (s\sqrt{2\omega}) & \cosh (s\sqrt{2\omega}) \ea\right)
	\left(\ba{c} \hx(0)+\omega^{-1}F\\
	\sqrt{\frac{2}{\omega}} \hp(0) \ea\right)\ ,
\ee
and the differential equation (\ref{de}) is explicitly written
as
\bea{explde}
	\align\frac{d}{ds}\la x''(s)|x'(0)\ra
	\nn
	 \align=\left\{i\left[
	F''\inv{2\om}\inv{\sinh^2(\sr)} F''
	+F'\inv{2\om}\frac{\cosh^2(\sr)}{\sinh^2(\sr)}F'
	-F''\frac{\cosh(\sr)}{\om\sinh^2(\sr)}F'\right]
	\right. \nn
	&&\ \ \ \ ~~~~\left.-isM^2(\phi(x'))
	-\tr\left(\sqrt{\frac{\om}{2}}\frac{\cosh(\sr)}{\sinh(\sr)}
	\right)\right\}\la x''(s)|x'(0)\ra\ ,
\eea
where we have expanded $M^2(\phi)$ around $x'$. This choice of
expansion point will not matter when we later take $x'=x''=x$.
The notation $F''$ and $F'$ means $F_\mu(x'')$ and $F_\mu(x')$.
The non--trivial functions of $\om$ should be
defined by power series expansions in $\om^\mu\,_\nu$
with suitable contractions with $F_\mu$.
The solution of \eq{explde}  in $d$--dimensional Minkowski space is
\bea{amp}
	\la x''(s)|x'(0)\ra &=&e^{-i\frac{\pi}{4}(d-2)}
	(4\pi s)^{-d/2} \det\left(\frac{\sr}{\sinh(\sr)}\right)^{1/2}
	\nn\align
	\exp\left\{i\left[-F''\frac{\coth(\sr)}{(2\om)^{3/2}}F''
	+F'\left(\frac{s}{2\om}-\frac{\coth(\sr)}{(2\om)^{3/2}}\right)F'
	\right.\right.\nn\align\left.\left.
	~~~~~~~~~~+F''\inv{\sqrt{2\om}}\inv{\om\sinh(\sr)}F'
	-sM^2\right]\right\}\ .
\eea
Finally, taking $x'=x''=x$ we find
\bea{finLag}
	\Lett&=&\frac{i}{2}\int_0^\infty\frac{ds}{s}(4\pi s)^{-d/2}
	e^{-i\frac{\pi}{4}(d-2)}
	\det\left(\frac{\sr}{\sinh(\sr)}\right)^{1/2}
	\nn
	&&\exp\left\{-is\left[M^2-F\inv{2\om}
	\left(1-\frac{2}{\sr}\tanh\frac{\sr}{2}\right)F\right]\right\}\ .
\eea
Notice that the integrand is well--defined as a power series
in $\om^\mu\,_\nu$ even when $\om^\mu\,_\nu$
is not invertible. The integrand has possible
poles on the real and imaginary axes, but only there since
eigenvalues of $\om$ are real.
We now want to deform the $s$--contour to the negative imaginary
axis. This is possible if the arc at infinity gives a vanishing
contribution. In our case we must require $M^2>\inv{2} F^\mu
(\om^{-1})_{\mu\nu}  F_\nu$. If we assume
 the eigenvalues of $\om_{\mu\nu}$ to be positive,
the condition on $M^2$ is the same as saying that the mass squared,
in the quadratic approximation
$M^2(x+\xi)\simeq M^2(x)-F_\mu\,
\xi^\mu+\inv{2}\xi^\mu\,\om_{\mu\nu}\,\xi^\nu$,
is positive everywhere, not only at $x$.
We now assume this condition to be satisfied and
perform the contour deformation.
If the above condition is not satisfied, we would expect an
imaginary part since the particles can then
 (in the quadratic approximation)
propagate to a point where they become tachyonic. Such an imaginary part
would be an artefact of the approximation rather than a physical
reality if the full $M^2(\phi)$ indeed is positive everywhere.

In 4 dimensions ($d=4$)
\eq{finLag} is infinite as it stands and has to be
renormalized. This can be done in a standard manner by adding
and subtracting
\be{ct}
	\inv{32\pi^2}(4\pi\mu^2)^{-\epsilon}
	\int_0^\infty\frac{ds}{s^{3-\epsilon}}
	\left(1+\frac{s^2}{6}\om^\alpha\,_\alpha\right)e^{-sM^2}\ .
\ee
It is not necessary
to renormalize
the term proportional to $\om^\alpha\,_\alpha$ since it is finite after a
partial integration
(the divergent part is a total derivative). We add it anyway
to make $\Lett$ finite before any partial integrations. The
regularized expression in $4-2\epsilon$ dimensions is then
\bea{L1ren}
	\Lett&=&\inv{32\pi^2}\int_0^\infty\frac{ds}{s^3}
	\left[\det\left(\frac{\sr}{\sin(\sr)}\right)^{1/2}
	\right.\nn
	&&\times\left.\exp\left\{sF\inv{2\om}
	\left(1-\frac{2}{\sr}\tan(\frac{\sr}{2})\right)F\right\}
	-\left(1+\frac{s^2}{6}\om^\alpha\,_\alpha\right)\right]
	e^{-sM^2}
	\nn\align
	+\frac{M^4}{64\pi^2}\left(\inv{\epsilon}+\frac{3}{2}
	-\gamma-\ln\frac{M^2}{4\pi\mu^2}\right)
	-\inv{32\pi^2}\om^\alpha\,_\alpha\ln\frac{M^2}{4\pi\mu^2}\ .
\eea
Even though \eq{L1ren} looks
manifestly real, there might be some imaginary contributions
from poles on the real $s$--axis, depending on the eigenvalues
of $\om$. The contour should go slightly above those poles.

At this point it is interesting to start to series--expand \eq{L1ren} in
derivatives to see if it coincides with
known results. We then find before any partial integration:
\bea{derexp}
	\Lett\simeq\align
	\frac{M^4}{64\pi^2}\left(\inv{\epsilon}+\frac{3}{2}
	-\gamma-\ln\frac{M^2}{4\pi\mu^2}\right)
	-\inv{32\pi^2}\frac{\om^\mu\,_\mu}{6}
	\ln\frac{M^2}{4\pi\mu^2}-\inv{32\pi^2}\frac{F^\mu F_\mu}{M^2}
	\nn[2mm]
	+\align
	\inv{16\pi^2}\left[\frac{(F_\mu F^\mu)^2}{96M^8}
	-\frac{F_\mu\om^\mu\,_\nu F^\nu}{60M^6}
	-\frac{F_\mu F^\mu\om^\nu\,_\nu}{72M^6}
	+\frac{\om^\mu\,_\mu\om^\nu\,_\nu}{144M^4}
	+\frac{\om^\nu\,_\mu\om^\mu\,_\nu}{180M^4}\right]\ .
\eea
The effective potential (no derivatives) and the wave
function renormalization (two derivatives) come
out correctly up to partial integration.
However, comparing with e.g.  \cite{Fraser85}
or \cite{MossTW92}  one finds an apparent discrepancy for the
four--derivative term (in \cite{MossTW92} there is also an
overall sign error). The difference is a total derivative
up to a term which contains a field with three derivatives:
\be{totder}
	\inv{180}\left(\frac{(\om^\mu\,_\mu)^2}{M^4}
	-\frac{2 F^\mu F_\mu \om^\nu\,_\nu}{M^6}\right)
	=\pa^\mu\left(\frac{F_\mu \om^\nu\,_\nu}
	{180 M^4}\right)-\frac{F_\mu \pa^\mu \om^\nu\,_\nu}
	{180 M^4}\ .
\ee
In the present approximation we neglect the last term
in \eq{totder} so that the
result here is actually consistent with \cite{Fraser85,MossTW92}.
This observation reminds us that the local effective Lagrangian
density in the effective action is not uniquely defined since
we can add total derivatives without changing the
equations of motion. Only the integral over space--time
or the equations of motion have intrinsic meaning.
%
\Section{zerofield}{Zero field limit for $m^2=0$}
At the end of Section \ref{Locexp} we saw what a typical
higher order term in the derivative expansion looks like
(see \eq{derexp}).
It is easy to see that, for dimensional reasons, the higher
the power of derivatives is, the higher the power of $M^2$ in
the denominator is. For a massless theory, $M^2=\lambda\phi^2/2$,
so for small $\phi$ this expansion obviously breaks down.
In the derivative expansion, corrections to derivative
terms are computed assuming that the field is constant.
Since the corrections
diverges we have to resum the series. After all, if $\pa_\mu\phi$
is non--zero then $\phi$ cannot be zero everywhere.
The propagator we calculated in Section \ref{Locexp} contains
the resummation necessary to make the $\phi\rightarrow 0 $
limit meaningful.
It does not really make sense to talk about $\phi=0$ in the
effective action since for instance $\pa_\mu\phi\pa^\mu\phi$
can be partially integrated to $\phi\pa_\mu\pa^\mu\phi$,
which would then be zero although it does contribute to the
equation of motion. In order to avoid the problems of possible
partial integration and mixing of derivatives we shall study
the equations of motion directly. From \eq{1ltp} we have
the unrenormalized equation of motion
\be{eqom}
	\frac{\delta\Gamma_{cl}}{\delta\phi}-
	\frac{\lambda\phi(x)}{2} \int_0^\infty ds
	\la x(s)|x(0)\ra = 0 \ .
\ee
It may at first look strange that the whole
one--loop contribution
in \eq{eqom} (using \eq{amp}) comes from
 varying \eq{EA} (using \eqs{1lLag}{finLag}) only
with respect to the $\phi$ in
$M^2$ and not in $F$ or $\omega$. But, rewriting \eq{EA} as
\be{EAxy}
	\Gamma_{cl}[\phi]-
	\frac{i}{2} \int_0^\infty\frac{ds}{s} e^{-isM^2(x)}
	\int d^4y\,\la y|\exp[is(\hp^2-M^2(\hat{y})+M^2(x))]|y\ra\ ,
\ee
one sees that the linear variation with respect
to $\phi$ only gets contributions from the first factor
$e^{-isM^2}$ and not from the operator
part in $M^2(\hat{y})$.
Partial integrations in
\eq{EA} would, of course, spoil this property.
Also the truncation of including only the second derivative,
as in \eq{finLag}, spoils this property in general.
We can verify explicitly from \eq{derexp} that up to
two derivatives
\be{twoder}
	\frac{\delta\Gamma}{\delta M^2}
	=\frac{\pa \cL}{\pa M^2}-\pa_\mu\frac{\pa \cL}{\pa F_\mu}
	+\pa_\nu\pa_\mu\frac{\pa\cL}{\pa\om_{\nu\mu}}
	=\frac{\pa \cL}{\pa M^2}\ ,
\ee
so that is does not matter whether it is only $M^2$ that
is varied with respect
to  $\phi$  or also $F$ and $\om$.
In
the case of four or more derivatives \eq{twoder} no longer holds
since for instance a term like $\pa^4 M^2/M^2$ would contribute
to the equations of motion
in the present approximation, but it is not included in
the effective Lagrangian in \eq{finLag}. This shows again
that one should discuss the equations of motion where
these ambiguities do not occur.

To simplify the analysis we shall now assume that
the background field is static and that the
eigenvalues of $\om_{ij}$ are positive.
This is natural since $M^2=\lambda\phi^2/2$ is positive for real
$\phi$ and has to be growing (or constant) in all directions
away from a point where $\phi=0$.
We continue to suppress the space--time indices
but in this section they should be understood as
only space indices with Euclidean metric.
Thus, we define $F$ and $\om$ by $F_i$ and $\om_{ij}$, with all indices
as subscripts, so there is a sign difference with $\om^\mu\,_\nu$.
In order to do a series expansion of \eq{eqom} in powers
of $\phi$ we write the one--loop part,
after renormalization, as
\bea{reneqom}
	\Gamma^{(1)}(x)=-\align
	\frac{\lambda\phi}{32\pi^2}\int_0^\infty \frac{ds}{s^2}
	\left[\det\left(\frac{\sr}{\sinh(\sr)}\right)^{1/2}
	\right.\nn\align\times\left.
	\left(\exp\left\{s\lambda^2\phi^2\pa_i\phi
	\left(\inv{2\om}\biggl(1-\frac{2}{\sr}\tanh(\frac{\sr}{2}
	)\biggr)\right)_{ij}
	\pa_j\phi\right\}-1\right) \right.\nn
	\align -\left. \left(1-\det\left(\frac{\sr}
	{\sinh(\sr)}\right)^{1/2}\right)\right]
	\exp\left(-s\frac{\lambda\phi^2}{2}\right)
	\nn
	\align +\frac{\lambda\phi}{32\pi^2}
	\frac{\lambda\phi^2}{2}\left(\inv{\epsilon}+1
	-\gamma-\ln\frac{\lambda\phi^2}{8\pi\mu^2}\right)\ .
\eea
The two terms in the square brackets are separately
finite for small $s$, and we shall treat them separately.
The first one can actually be expanded directly in $\phi$
since the determinant makes the integral convergent for
large $s$.
The second term has to be treated with more care.
One way to evaluate it is to divide the $s$--integral
into two intervals, $[0,\Lambda]$ and $[\Lambda,\infty]$,
where $ \lambda \phi^2\ll \Lambda^{-1}\ll\norm{\om}^{1/2}$.
Doing suitable
approximations and partial integrations on each interval, we
finally obtain
\bea{expeqom}
	\Gamma^{(1)}(x)\simeq-\frac{\lambda\phi}{32\pi^2}\align
	\left\{\lambda^2\phi^2
	\int_0^\infty\frac{ds}{s} g(s)\, \pa_i\phi
	\left[\inv{2\om}\left(1-\frac{2}{\sr}\tanh
	(\frac{\sr}{2})\right)\right]_{ij}\pa_j\phi
	\right.\nn\align\!\!\!\!\!\left.
	-\int_0^\infty\frac{ds}{s^2}(1-g(s))
	+\frac{\lambda\phi^2}{2}\left(-\inv{\epsilon}
	+\int_0^\infty ds \ln(4\pi\mu^2s)\frac{dg(s)}{ds}
	\right)\right\}\ ,
\eea
where
\be{gdef}
	g(s)\equiv\det\left(\frac{\sr}{\sinh(\sr)}\right)^{1/2}\ .
\ee
It is interesting to see that \eq{expeqom} has a well--defined
and finite expansion in powers of $\phi$.
The reason why it is IR--finite, i.e. why it is
convergent for large $s$, is that $g(s)\rightarrow 0$
rapidly.  We do not see immediately from the treatment
here how higher order derivatives of $M^2$ would affect the
result, but we can expect the correction to be finite
on physical grounds if there are no instabilities.
%
\Section{finT}{Finite temperature effective action}
There are two main formalisms for doing finite temperature
field theory, the real--time and the imaginary--time
formalisms, and both can be used to generalize the
calculations in Section \ref{Locexp}. We shall discuss
both formalisms here and use them for different purposes.
\\[10mm]\ \
{\large\sc Real--time formalism}
\\[4mm]
The generalization to the real--time formalism is done
by using the thermal propagator when
calculating the tadpole in \eq{1ltp}. Although a doubling of the number of
degrees of freedom is generally needed for a consistent loop
expansion, only the (11)--part of the propagator is needed
to calculate the tadpole at one loop. The thermal (11)--propagator
is given by
\be{thprop}
	\frac{i}{p^2-M^2+i\epsilon}
	+2\,f_B(p_0)\,\Re\frac{i}{p^2-M^2+i\epsilon}\ .
\ee
The real part can be represented by extending the $s$--integral
in \eq{1lLag} from $-\infty$ to $\infty$, but the
phase of the normalization in
\eq{amp} is only valid for $s>0$. Instead of
keeping track of the phase factor
we continue with explicitly taking the real part.
The order of $f_B(p_0)$ and $M^2(\phi(\hx))$ in \eq{thprop}
is not well--defined,
unless the background field is time--independent. It also
makes sense to limit the considerations to such fields since the
system will not remain in equilibrium otherwise. It is
conceivable to treat small time--dependent perturbations
of an equilibrium background, assuming the perturbation to
be small, but here we also want to deal with large
field amplitudes. Thus, assuming $F_0=0$ and
$\om_{\mu 0}=0$, we get
\bea{thLag}
	\cL^{\beta,\mu}(\phi)&=& - \Re\int \frac{dp_0}{2\pi}
	f_B(p_0)e^{-i\frac{\pi}{4}}\int_0^\infty \frac{ds}{s}
	(4\pi s)^{-3/2} \nn
	\align\times\left[\det\left(\frac{\sr}{\sin(\sr)}\right)^{1/2}
	\exp\left[iF\frac{s}{2\om}
	\left(1-\frac{2}{\sr}\tan(\frac{\sr}{2})\right)F
	\right]-1\right]\nn
	\align\times\exp\left[is(p_0^2-M^2)\right]\nn
	\align +\inv{12\pi^2}\int dp_0 f_B(p_0)
	(p_0^2-M^2)^{3/2}\Theta(p_0^2-M^2)\ .
\eea
The last term in \eq{thLag} is the effective potential, which
comes out when we regularize for small $s$ and use a
$\Gamma$--function to define a finite value.

This formal expression of the thermal effective action is
rather difficult to use in general, since we would like
to deform the $s$--contour to the positive imaginary axis
for large $p_0$, but then we encounter all the poles
on the positive real axis (the $s$--contour as it stands goes slightly
below the real axis). At least if we assume the eigenvalues
of $\om_{ij}$ to be positive, which we want for stability.
The situation is reminiscent of electrons in a
constant magnetic field \cite{ElmforsPS94} where the poles
correspond to Landau levels, while here they are related to
oscillations in a harmonic oscillator potential.

As in Section \ref{Locexp}, we want to compare \eq{thLag} with
known derivative expansions. The effective potential is
easily recognized to be correct from the last term in \eq{thLag}.
The wave function renormalization
comes from two terms, $\om_{ii}$ and $F_i F_i$.
A naive expansion in $F_iF_i$ of
the exponent in \eq{thLag} leads to a divergent integral,
but if one first makes a partial integration with respect
to $p_0$ it becomes finite. In the end the total contribution
agrees with the high temperature expansion in \cite{MossTW92}.
\\[10mm]
{\large\sc Imaginary--time formalism}
\\[4mm]
The transition to the imaginary--time formalism is most easily
performed in a $p_0$ rather than $x_0$ representation of
the amplitude $\la x''(s)|x'(0)\ra$. In the limit of coinciding
points we can write
\be{itfamp}
	\la x(s)|x(0) \ra =
	\int \frac{dp_0}{2\pi} dx'_0 \la x'_0(s),x_i(s)|x_0(0),x(0)_i \ra
	e^{ip_0(x'_0-x_0)} \ ,
\ee
and replace $\int dp_0$ by $i2\pi T\sum_n (p_0\goto i2\pi T n)$, as usual.
Let us now study the equation of motion and see what happens
in the $\phi\goto 0$ limit when $m^2=0$.\footnote{
To be consistent in the high temperature limit we would have to
resum the thermal mass correction $\lambda T^2/24$.
A  zero effective mass is found at a second order phase transition
in a spontaneously broken theory where the resummed mass is
$m^2(T)=-\mu^2+\lambda T^2/24$, $\mu^2>0$. This is really the
situation we imagine when we write $m^2=0$ here.
}
The full one--loop correction to the
equation of motion is
\bea{itfeqom}
	\Gamma^{(1)}[\phi]\align+\Gamma^{(1)}_\beta[\phi] =
	-\frac{\lambda\phi}{2}\int_0^\infty
	ds\, e^{-i\frac{\pi}{4}}T\sum_{n=-\infty}^\infty
	(4\pi s)^{-3/2} \det\left(\frac{\sr}{\sin(\sr)}\right)^{1/2}
	\nn
	\align \times\exp\left[-is\left( (2\pi T n)^2+M^2
	-F\frac{s}{2\om}
	\left(1-\frac{2}{\sr}\tan(\frac{\sr}{2})\right)F\right)\right]\ .
\eea
We shall only study the leading contribution for small $\phi$,
so we can put $M^2=F=0$. The IR convergence is anyway governed
by $\om$. The zero temperature part of \eq{itfeqom} can be
extracted after a Poisson resummation, and in the remaining
piece we deform the contour by taking $s\goto -is$. Thus the finite
value of $\Gamma^{(1)}_\beta$ for  small $\phi$
is
\be{thGlim}
	\Gamma^{(1)}_\beta[\phi]\simeq
	-\frac{\lambda\phi}{16\pi^2}\int_0^\infty\frac{ds}{s^2}
	\det\left(\frac{\sr}{\sinh(\sr)}\right)^{1/2}
	\sum_{n=1}^\infty\exp\left(\frac{-n^2}{4T^2s}\right) \ .
\ee
To get some more concrete information out of \eq{thGlim} we
can take the high temperature limit where the determinant
can be approximated by $1$. As expected we find
$\Gamma^{(1)}_\beta =- \lambda T^2\phi/24$, which simply is the thermal
mass shift. Therefore, we also want the next to leading term
in order to get something new. Again, a straightforward
expansion of the determinant does not work since it gives
divergent remaining integrals. This is not surprising since
it is the large--$s$ behaviour of  the determinant that
should make the expansion IR--convergent. One way to get around
the problem is to first subtract the 1 from the
determinant and then use the formula
\be{vikformel}
	\sum_{n=1}^\infty e^{-n^2/s}=
	\inv{2}(\sqrt{\pi s}-1)+\sqrt{\pi s}
	\sum_{n=1}^\infty e^{-n^2\pi^2 s}\ ,
\ee
and a transformation $s\goto 1/s$ to restrict the integration
interval to $[1,\infty]$. We then obtain for the subleading
contribution in $T$
\bea{vikeqom}
	\Gamma^{(1)}_\beta[\phi]+\frac{\lambda T^2\phi}{24}=\align
	-\frac{\lambda\phi T^2}{4\pi}
	\int_1^\infty ds \left[ \left(g(\inv{4\pi T^2 s})-1\right)
	+s^{-3/2}\left(g(\frac{s}{4\pi T^2})-1\right)\right]
	\sum_{n=1}^\infty e^{-n^2\pi s}
	\nn
	\align-\frac{\lambda\phi T^2}{8\pi^2} \int_1^\infty ds
	\left(g(\frac{s}{4\pi T^2})-1\right)(s^{-3/2}-s^{-2}) \ ,
\eea
using  the notation in \eq{gdef}.
The first term on the right--hand side of \eq{vikeqom}
goes to zero in the high temperature limit, but
the second term  finally gives:
\be{hTeqom}
	\Gamma^{(1)}_\beta[\phi]\simeq-\frac{\lambda T^2}{24}\phi
	-\frac{\lambda T}{16\pi^{3/2}} \phi
	\int_0^\infty \frac{ds}{s^{3/2}}
	\left[\det\left(\frac{\sr}{\sinh(\sr)}\right)^{1/2}-1\right]\ .
\ee
We can then conclude that also the finite temperature
part of the equation of motion is well--behaved
when $\phi\goto 0$ as long as $\om$ has positive eigenvalues.
%

\end{document}